\documentclass[english]{iopart}
\usepackage[T1]{fontenc}
\usepackage[latin9]{inputenc}
\usepackage{verbatim}
\usepackage{float}
\usepackage{amssymb}
\usepackage{graphicx}

\makeatletter

\floatstyle{ruled}
\newfloat{algorithm}{tbp}{loa}
\providecommand{\algorithmname}{Algorithm}
\floatname{algorithm}{\protect\algorithmname}

\@ifundefined{definecolor}
 {\usepackage{color}}{}
\@ifundefined{definecolor}
 {\@ifundefined{definecolor}
 {\usepackage{color}}{}
}{}
\usepackage{babel}

\usepackage{babel}\usepackage{amsfonts}\@ifundefined{definecolor}
 {\@ifundefined{definecolor}
 {\usepackage{color}}{}
}{}
\usepackage{array}\usepackage{supertabular}\usepackage{hhline}

\newcommand{\arraybslash}{\let\\\@arraycr}

\newcommand{\ps@Standard}{
  \renewcommand\@oddhead{}
  \renewcommand\@evenhead{}
  \renewcommand\@oddfoot{}
  \renewcommand\@evenfoot{}
  \renewcommand\thepage{\arabic{page}}
}

\usepackage{babel}

\newcommand{\ket}[1]{|#1 \rangle}

\newcommand{\sandwich}[3]{\left \langle #1 \mid #2 \mid #3 \right\rangle}

\@ifundefined{showcaptionsetup}{}{%
 \PassOptionsToPackage{caption=false}{subfig}}


\usepackage{babel}
\usepackage[normalem]{ulem}

\usepackage{algorithm,algpseudocode}

\makeatother

\usepackage{babel}

\begin{document}

\title{Modernizing Quantum Annealing using Local Searches}

\author{Nicholas Chancellor\footnote{email: nicholas.chancellor@durham.ac.uk}}

\address{Department of Physics, Durham University, South Road, Durham, UK}
\begin{abstract}
I describe how real quantum annealers may be used to perform local
(in state space) searches around specified states, rather than the
global searches traditionally implemented in the quantum annealing
algorithm. Such protocols will have numerous advantages over simple
quantum annealing. By using such searches the effect of problem mis-specification
can be reduced, as only energy differences between the searched states
will be relevant. The quantum annealing algorithm is an analogue of
simulated annealing, a classical numerical technique which has now been superseded.
Hence, I explore two strategies to use an annealer in a
way which takes advantage of modern classical optimization algorithms.
Specifically, I show how sequential calls to quantum annealers can
be used to construct analogues of population annealing and parallel
tempering which use quantum searches as subroutines. The techniques
given here can be applied not only to optimization, but also to sampling.
I examine the feasibility of these protocols on real devices and note
that implementing such protocols should require minimal if any change
to the current design of the flux qubit-based annealers by D-Wave
Systems Inc. I further provide proof-of-principle numerical 
experiments based on quantum Monte Carlo that
demonstrate simple examples of the discussed techniques.

\end{abstract}

\maketitle

\tableofcontents
\markboth{Modernizing Quantum Annealing using Local Searches}{Modernizing Quantum Annealing using Local Searches}

\section*{\textcolor{red}{note added after publication}}
I have been made aware of a previously existing related work which relates to incorporating guesses into a closed system adiabatic protocol:

A. Perdomo-Ortiz, S. E. Venegas-Andraca, \& A. Aspuru-Guzik Quantum Inf Process (2011) 10: 33. doi:10.1007/s11128-010-0168-z \cite{Perdomo-Ortiz2011}

While the protocol proposed in this work is actually quite different then the one I propose, the idea is related enough that it should be cited.  I have also been made aware of the following two machine learning papers, which should have been cited in the introduction:

 M. Benedetti, J. Realpe--G{\'o}mez, R. Biswas,
 and A. Perdomo-Ortiz, Estimation of effective temperatures in quantum annealers for sampling applications: A case study with possible applications in deep learning. Phys. Rev. A, 94:022308, Aug 2016. \cite{Benedetti2016}

and

M. Benedetti, J. Realpe--G{\'o}mez, R. Biswas, and A. Perdomo-Ortiz. Quantum-assisted learning of graphical models with arbitrary pairwise connectivity. arXiv:1609.02542, 2016. \cite{Benedetti2016a}

\section{Introduction}

Recently, there has been much interest in using the Quantum Annealing
Algorithm (QAA) \cite{Finella1994,Farhi2001,Brooke1999} which utilizes
quantum tunneling to aid in solving commercially interesting problems.
A complete list of all potential applications would be too long to
give here. However applications have been studied in such diverse
fields as finance \cite{Marzec}, computer science \cite{Choi2010},
machine learning \cite{Adachi2015,Denil,Rose(2014),Amin2016}, communications
\cite{Chancellor2015,Otsubo(2012),Otsubo(2014),Jordan(2006)}, graph
theory \cite{Vinci(2014)}, and aeronautics \cite{Coxson(2014)},
illustrating the importance of such algorithms to real world problems.
While some of these applications rely on the ability of the QAA to
perform optimization by finding the lowest energy state of a classical
problem Hamiltonian, others such as \cite{Adachi2015,Denil,Rose(2014),Amin2016,Chancellor2015,Jordan(2006)},
instead rely on the fact that open quantum systems effects allow for
sampling of an approximate Boltzmann distribution. I will discuss
both of these techniques in due course. 

The archetypal model for quantum annealing, because of its connection
to condensed matter physics as well as the fact that it can be implemented
on real devices \cite{D-Wave} is the transverse field Ising model,
with Hamiltonian $H(s)$ given by

\begin{equation}
H(s)=-A(s)\,\sum_{i}\sigma_{i}^{x}+B(s)\,H_{Problem},\label{eq:AnnealerHam}
\end{equation}
where

\begin{equation}
H_{Problem}=-\sum_{i}h_{i}\sigma_{i}^{z}-\sum_{i,j\in\chi}J_{ij}\sigma_{i}^{z}\sigma_{j}^{z}.\label{eq:ISGham}
\end{equation}
encodes the problem of interest, $\chi$ is the hardware graph, and
$A(s)$ and $B(s)$ are the annealing schedule, which determines how
the energy scales of the transverse and longitudinal terms change
with the annealing parameter, $s\in\left(0,1\right)$. The problem
is encoded by speficifying the values of $h_{i}$ and $J_{ij}$. For
the QAA, $A(0)\gg B(0)$ and $A(1)\ll B(1)$, and $A(s)$ decreases
monotonically while $B(s)$ increases monotonically with increasing
$s$. Applying the QAA consists of monotonically increasing $s$ with
time such that the ground state of the system changes over time between
the (known) ground state of the transverse part of the Hamiltonian
($\sum_{i}\sigma_{i}^{x}$) to the solution of the (classical) problem
to be solved, Eq. (\ref{eq:ISGham}). The search space of the transverse
Ising model is a hypercube where each vertex corresponds to a bitstring,
the dimension is equal to the number of qubits, and the Hamming
distance between classical states corresponds to the number of edges
which must be traversed between the states. This structure is independent
of the interaction graph defined by $J_{ij}$ which along with   $h_{i}$
determine the energy at each vertex.

I choose to focus on the transverse field Ising model for concreteness,
and because the action of the transverse field is a quantum analogue
of single bit flip updates in classical Monte Carlo. However, the
arguments presented in this paper should hold for most other search
spaces as well, with Hamming distance replaced with a more general
notion of search space distance.

The QAA can be though of as analogous to classical Simulated Annealing
(SA) in which quantum fluctuations mediated by the addition of non-commuting
terms to a classical Hamiltonian, play the role which temperature
plays in SA. Simple SA, however, has been superseded by more sophisticated
algorithms, such as parallel tempering \cite{Swendsen1986,Earl2005},
population annealing \cite{Hukushima2003,Matcha2010,Wang2015}, and
isoenergetic cluster updates \cite{Zhu2015} to name a few. This then
begs the question of whether quantum annealing hardware can be used
in a clever way to gain the advantages of these modern classical algorithms,
by using a hybrid algorithm employing both quantum and classical search
techniques, or by using multiple quantum searches in a sequential
way to make algorithmic gains. 

The QAA, as it is currently designed, is not amenable to such adaptations.
It is a global search, and there is no obvious way to insert information,
from either a classical algorithm or previous runs of the QAA, in
a meaningful way to improve the performance. Furthermore, the QAA
is fundamentally different from classical annealing in that, due to
the famous no-cloning theorem \cite{Wooters(1982)} of quantum mechanics,
we cannot determine exactly what the intermediate state of the system
is part way though the anneal. This is in direct contrast to SA, where
every intermediate state is known, and can be manipulated arbitrarily
to build better algorithms. For example, classical gains can be made
by running many runs in parallel and probabilistically replacing poor
performing copies with those which are performing well (population
annealing), or raising the temperature for those which perform poorly
and lowering it for those which perform well (parallel tempering). 

In order to build quantum versions, let us consider a subroutine similar
to QAA, but which performs a local search of a region of phase space
with a controllable size around a user selected initial state. The
input and output of a single step of this algorithm is completely
classical, so the no-cloning theorem is no longer a barrier and these
local quantum searches can be combined arbitrarily with both other
quantum searches and classical searches. Using this, I construct analogues
to state-of-the-art classical algorithms, but made of quantum building
blocks, I also demonstrate how to construct new hybrid algorithm which
can use any classical algorithm which meets a very general set of
criteria as a subroutine. It is worth pointing out here that this kind of
search has been considered in a limited scope in recent work by others \cite{Amin_patent}.

I further argue %
that these subroutines will be less sensitive to noise in the form
of problem mis-specification than the QAA. The typical random energy
differences between states due to these errors scales like $\sqrt{N}$,
where $N$ is the total number of qubits. A local search, however,
only searches a small subspace of the total solution space, and therefore
only errors which occur on states within this subspace are relevant.
A local search can therefore give meaningful results even in a problem
where the global optimum is no longer correctly specified due to noise.

This manuscript is structured as follows. In Sec.~\ref{sec:Optimization-and-Sampling}
I give a brief overview of the currently used optimization and sampling
techniques which will be discussed in this manuscript. In Sec.~\ref{sec:Hybrid-Computing}
I will explain how a hybrid technique can be constructed by combining
local quantum searches with local classical searches. Following that
section, in Sec.~\ref{sec:How-much-to} I examine the ways in which
adaptive search ranges can be used, including construction of analogues
of parallel tempering and population annealing. In Sec.~\ref{sec:Local_search}
I describe how a local search can be achieved using an annealer which
implements a transverse field Ising model.  I next perform simple numerical proof of principle experiments using Quantum Monte Carlo (QMC) techniques in Sec.~\ref{sec:proof_of_principle} on the simplest of these algorithms to demonstrate the value of local searches. Next, I discuss how these methods
can be extended to sampling applications in Sec.~\ref{sec:Applications-to-Sampling}
and describe why local searches should be more robust against problem
mis-specification in Sec.~\ref{sec:How-to-avoid}. Finally in Sec.~\ref{sec:Hardware-Implementation}
I examine the feasibility of implementing such a protocol on real
devices, and conclude with general discussion in Sec.~\ref{sec:Conclusions}.

\section{Optimization and Sampling Techniques\label{sec:Optimization-and-Sampling}}

To explain the new method I propose, it is useful to first summarize
some classical and quantum optimization techniques which are currently
studied. The list given here is not intended to be exhaustive, and
in particular will only cover some of the classical techniques from
the Monte Carlo `family' of methods, those which use Metropolis weighted
updates. A Metropolis update is an update which is performed probabilistically
in the following way: if the update lowers the energy of a state it
will be performed with $100\%$ probability; however if the update
increases the energy than it will be performed with a probability
equal to $\exp(-\frac{\Delta E}{T})$ where $\Delta E$ is the change
in energy and $T$ is an effective temperature. Because Metropolis
updates obey detailed balance, they can be used to sample a thermal
distribution as well as to find ground states, assuming all other
update steps also obey detailed balance.

\subsubsection*{Simulated Annealing (SA)}

In SA an initial state is chosen randomly and is updated by flipping
individual spins according to Metropolis rules. The temperature parameter
is then reduced according to an annealing schedule until $T=0$ is
reached. SA derives its name from the fact that Metropolis rules updates
obey detailed balance, and therefore an SA run can be thought of as
a simulation of a physical spin system which is cooled under classical
dynamics.

\subsubsection*{Parallel Tempering}

In parallel tempering \cite{Swendsen1986,Earl2005}, multiple copies
of a system are initialized in random states each with a different
temperature parameter. Spin flip Metropolis updates are applied on
each copy. These temperature parameters are kept fixed, but additional
update rules are applied to swap the temperature of copies probabilistically
in a way which obeys detailed balance. These rules mean that poorly
performing copies have a high probability of having their temperature
raised so that they perform a long range exploration of the solution
space, whereas the temperature of copies which perform well are reduced
so that their search becomes more local. Because parallel tempering
is able to abort explorations of regions for which the algorithm performs
poorly it provides a substantial improvement over SA. There is no
known physical phenomenon which is an analogue of parallel tempering.

\subsubsection*{Population Annealing}

Population annealing \cite{Hukushima2003,Matcha2010,Wang2015} also
uses multiple copies of the same system, again initialized in random
states. Unlike what is done in parallel tempering, spin flip Metropolis
updates are performed on all copies at the same temperature and it
is slowly reduced according to an annealing schedule. In population
annealing there is also an update rule beyond simple Metropolis updates.
These rules probabilistically delete poorly performing copies, and
replicate those which perform well in a way which not only obeys detailed
balance but also for which the average population remains constant
and does not explode exponentially, or decay to zero. As with parallel
tempering, population annealing contains a mechanism to abort searches
which perform poorly, and provides a substantial improvement over
SA. The performance gains over SA from parallel tempering and population
annealing have been observed to be comparable \cite{Hukushima2003,Matcha2010,Wang2015}.
Again, there is no known physical phenomenon which is an analogue
to population annealing.

\subsubsection*{Quantum Annealing Algorithm (QAA)}

The QAA is an algorithm in which quantum fluctuations act in an analogous
way to the way in which Metropolis updates are used in SA. The strength of the quantum
fluctuations is then slowly turned down according to an annealing
schedule, this could be performed for instance using the transverse
field Ising model Hamiltonian in Eq.~\ref{eq:AnnealerHam}. For the
purposes of this paper, I use QAA to refer to the process by which
quantum fluctuations are slowly turned down, and not the specific
nature of the fluctuations. There are actually two mechanisms by which
the QAA can obtain the ground state of the Hamiltonian. In a closed
quantum system setting the adiabatic theorem of quantum mechanics
guarantees that, for slow enough evolution, the system will remain
in its ground state as long as it is initialized in its ground state \cite{Ehrenfest1916,Born1928,Cheung2011},
this is conventionally called Adiabatic Quantum Computation (AQC) \cite{Farhi2001}.
For discussion of the effect of open quantum system phenomena on AQC,
see \cite{Venuti2015}. On the other hand, in an open quantum system
setting with a low temperature bath, interactions with the bath can
lead transitions toward lower energy states, this mechanism is known
as Quantum Annealing (QA) \cite{Finella1994}. The QAA is performed
on an analog physical system, which can be thought of as an analog
computer. Rather than being an analogue to a physical process as SA
is, the QAA is a physical process itself. QAA-like protocols have
also been successfully implemented in condensed matter systems \cite{Brooke1999}.
While the quantum distribution obtained from the QAA will not generally
be the thermal distribution one would find with zero transverse field,
it has been demonstrated experimentally that a quantum annealer can
be used for thermal sampling under some circumstances \cite{Chancellor2015},
it has also been demonstrated numerically that in some cases the transverse
field can act as an effective proxy for finite temperature \cite{Otsubo(2012),Otsubo(2014)}.
Recent work has also suggested that the dissipation from open quantum
system effects in QA can lead to an improvement in performance over
AQC \cite{Smelyanskiy(2015)}.

\subsubsection*{Quantum Random Walk }

A continuous time quantum random walk \cite{Farhi(1998)} is a quantum
protocol by which a quantum system is allowed to evolve under a fixed
quantum Hamiltonian to explore a solution space. It has been demonstrated
that continuous time quantum random walks yield a similar degree of
quantum advantage as other quantum search algorithms \cite{Shenvi2003},
for further extensions to spatial searches, see \cite{Childs2004}.
As I discuss later, in the limit of instantaneous annealing, the local
search protocol which I propose is a quantum random walk with a localized
starting condition and subject to noise from a fixed temperature bath.

\subsubsection*{Comparison Between Techniques}

Whether the dominant search mechanism is open system effects (QA),
or the adiabatic theorem of quantum mechanics (AQC), the QAA uses
a single monotonic anneal, and therefore is a quantum analog of SA.
There is no sense in which the QAA can abort poorly performing searches
in favor of those which perform better. The reason that using the
QAA may still be preferred over either parallel tempering or population
annealing is that QAA may grant a large quantum advantage. If such
an advantage were larger than the classical advantage which advanced
techniques such as parallel tempering or population annealing have
over SA, then the QAA would be preferable. However, it is possible,
especially for early generations of annealers, that only a modest
advantage, which is not as large as the advantage from advanced classical
techniques is present. In this case the QAA would still be out-performed
by these advanced techniques so would not be a desirable approach
to real world problems. However a technique which could gain a small
quantum advantage (possibly not even as large as the one which the
QAA has over SA) along with the advantages of advanced classical techniques
would still represent an improvement in the state-of-the-art. I examine
three possible approaches to achieve this goal: a hybrid algorithm
which combines local quantum searches with any classical algorithm
which follows a very general set of criteria (Sec.~\ref{sec:Hybrid-Computing}),
an analogue of parallel tempering which uses a local quantum search
as a subroutine (Sec.~\ref{sec:How-much-to}), and an analogue of
population annealing which uses a local quantum search as a subroutine
(Sec.~\ref{sec:How-much-to}).

\section{Hybrid Computing\label{sec:Hybrid-Computing}}

It has been demonstrated experimentally that the QAA performs well
in problems characterized by tall thin barriers in the energy landscape
\cite{Denchev(2016)}. On the other hand, from elementary arguments
about quantum tunneling amplitudes \cite{LeBallac_book} it is expected
to perform relatively poorly for energy landscapes characterized by
wide, flat barriers. While it is possible to experimentally generate
problems which are characterized by an energy landscape consisting
of thin barriers \cite{Boixo2016,Boixo2014} it is unclear whether
any problems with this structure are interesting in the real world.
While looking for such problems is one interesting route for research,
it is also worth thinking about how annealers may yield a benefit
in phase spaces characterized by a mixed energy landscape, with many
barriers of varying width.

\begin{figure}
\begin{centering}
\includegraphics[width=7cm]{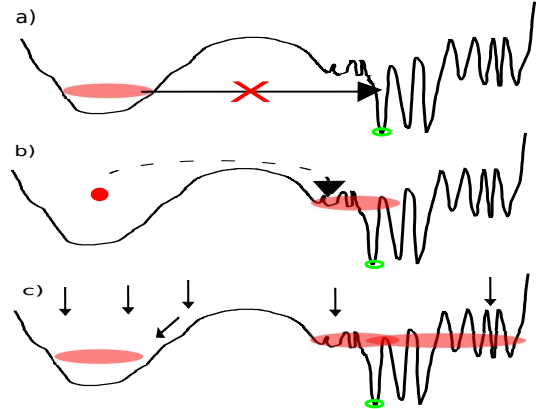}
\par\end{centering}

\caption{\label{fig:landscape_cartoon}Schematic representation way in which
a hybrid algorithm can potentially gain an advantage in finding the
local energy minima (circled in green) a) Wide energy barrier blocks
tunneling for the QAA, leading to suboptimal solution in smooth region
of the energy landscape. b) A classical algorithm can be used to overcome
the `smooth' features of the energy landscape, after which quantum
tunneling can be used to explore rough parts of the landscape. c)
In this hypothetical example even random initialization can help the
search if the initial wide energy barrier is avoided by chance. }
\end{figure}

In general one would not expect the phase space of a real optimization
problem to have only thin energy barriers for which quantum tunneling
yields a major benefit. One advantage of using local searches is that
different methods can be used to traverse different types of features
of the landscape. As an example, imagine the global optimum lies in
a `rough' region of phase space with many tall thin energy barriers,
but the phase space also has some relatively `smooth' global structure
which would require tunneling through a very wide barrier as shown
in Fig.~\ref{fig:landscape_cartoon}. In such a case the traditional
QAA would perform poorly. On the other hand a classical search should
be able to efficiently explore the larger features of the space and
could easily find the rough region of phase space where quantum tunneling
could then provide a major advantage. 

One key point about this example is that a quantum search would benefit
even from \emph{random} initialization, tunneling probability drops
exponentially with barrier width, so if the barrier in the `smooth'
part of the space was wide enough, the probability of successful tunneling
to the rough region in the QAA would be zero for all practical purposes.
On the other hand, if we assume that the quantum algorithm explores
the rough region efficiently, then on random initialization the probability
of success will be approximately proportional to the phase space volume
of the rough region.

In a more complex phase space, optimization may potentially benefit
from sequential use of a variety of algorithms in stages, feeding
results between various classical algorithms and the quantum device.
Any classical algorithm or the QAA could be used to initialize such
a protocol, but intermediate stages would have to be able to take
one or more input value and search based on that input. Fortunately,
most classical algorithms are structured in such a way that they store
intermediate states which are updated sequentially and therefore meet
these criteria. The entire Monte Carlo `family' of algorithms (parallel
tempering, population annealing, isoenergetic cluster updates etc...),
for instance could all be used this way, as could genetic algorithms.
As we discuss in Sec. \ref{sec:How-much-to}, analogues of many of
these classical Monte Carlo based protocols can be constructed and
implemented with local search on a quantum annealer.

As I discuss in the next section, the important characterizing feature
of a local search is its range. In the next section I discuss how
the ability to control the range of a search can be used to construct
hybrid quantum search algorithms. In contrast to the methods discussed
in this section, where individual calls to a local quantum search
can be interspersed with local classical search algorithms, in the
methods I discuss in that section, all searches will be performed
quantum mechanically, subject to an overall classical control structure
which decides which local searches will be performed and at which
range at each subsequent step.

\section{How much to Search\label{sec:How-much-to}}

Let us now examine more closely how control of the range of a local
search may be used to gain a computational advantage. In the previous
section I have outlined the basic idea of constructing hybrid algorithms
from classical and quantum local searches, but have not yet specified
how control over search range can be utilized. For this section, I
will assume the existence of a local search protocol which searches
a region around a point in solution space where the range is defined
by a parameter $s'$, with $s'=1$ corresponding to no search being
performed (stays in initial state with $100\%$ probability), and
$s'=0$ corresponding to a global search as performed by the traditional
QAA, and intermediate values correspond to intermediate ranges. In
Sec.~\ref{sec:Local_search} I will discuss how to implement such
a protocol on an annealer an abstract way using the traverse field
Ising model and demonstrate that such a local search can be accomplished
in a schedule which anneals to the point $A(s=s')$, $B(s=s')$ after
a preparation protocol. In Sec.~\ref{sec:Hardware-Implementation}
I will discuss the possibility of implementing such a search on real
devices. 

I will not assume that the functional dependence of the range of the
search in terms of the mean Hamming distance explored, $\mathfrak{h}(s')$
is known, only that the range monotonically increases with decreasing
$s'$. I further will assume that $\mathfrak{h}(s')$ may vary depending
on the Hamiltonian or initial state chosen. One could choose $s'$
heuristically either by finding a value of $s'$ which \emph{typically}
explores within a desired range for a given class of Hamiltonians,
or by doing several runs each time for different values $s'$ of the
local search subroutine is called and outputting the overall best
solution(s) found. While these heuristic techniques should work in
principle, they are probably sub-optimal, so we therefore discuss
a more sophisticated ways to approach this problem. Additionally we
discuss how this freedom to choose $s'$ can be used to build analogues
of powerful classical algorithms, but which make use of a quantum
processor.

I assume that each call to the annealer, which we will refer to as
an \emph{annealing run} and as the function ANNEALER\_CALL in our
algorithms actually consists of multiple searches around the same
point and defined by the same parameter, $s'$ each of which I refer
to as an individual \emph{annealing cycle}.

If a specific search radius is desired, for example by a maximum amount
of error which can be tolerated due to problem mis-specification,
then this search can be done adaptively. This is possible because
while it may be extremely difficult to \emph{predict} how much phase
space will be explored in a given run, it is easy to check \emph{experimentally.
}All one needs to do is to gather a statistically significant sample
and check the typical Hamming distance from the initial state. Based
on these distances, $s'$ can be either reduced or increased. If the
bisection adaptive search protocol given in algorithm \ref{alg:adaptive_s}
is used, the number of runs required scales logarithmically with the
desired accuracy of $s'$. Logarithmic scaling is achieved because
the bisection method halves the search range for $s'$ at each step,
therefore, assuming that the search range has been estimated correctly,
the precision of the parameter $s'$ will improve exponentially at
each step.

\begin{algorithm*}
\begin{algorithmic}[1]
\Procedure{adaptive\_s\_prime}{$H,state,dist,Nstep$} \Comment{adaptive procedure for finding s' which searches a specified volume of phase space}
	\State $s\_prime \gets 1/2$
	\State $s\_min \gets 0$
	\State $s\_max \gets 1$
	\For{$i=1,i++,Nstep$}
		\State $results \gets \textrm{ANNEALER}\_\textrm{CALL}(H,state,s\_prime)$ \Comment{Call to quantum annealer}
		\State $run\_dist \gets \textrm{TYPICAL}\_\textrm{HAMMING}\_\textrm{DIST}(results)$ \Comment{Get typical Hamming distance}
		\If{$run\_dist<dist$}
			\State $s\_max \gets s\_prime$
			\State $s\_prime \gets s\_min+0.5*(s\_prime-s\_min)$     \Comment{halfway between $s\_min$ and $s\_prime$}
		\Else
			\State $s\_min \gets s\_prime$
			\State $s\_prime \gets s\_max+0.5*(s\_max-s\_prime)$  \Comment{halfway between $s\_max$ and $s\_prime$}
		\EndIf
	\EndFor
\State \Return $s\_prime$
\EndProcedure
\end{algorithmic}

\caption{\label{alg:adaptive_s}Adaptive determination of $s'$ to explore
a region of phase space with a given size, see Sec. \ref{sec:Local_search}
for the details of how the local search given in ANNEALER\_CALL can
be implemented.}
\end{algorithm*}

Another method is to use the fact that the strength of the transverse
field, and by extension $s'$, can act as a proxy for temperature
to create analogues of familiar classical algorithms, which use quantum
rather then thermal fluctuations to compute. I will examine here ways
to create analogues of two such classical algorithms, parallel tempering
and population annealing. 

\begin{figure}
\begin{centering}
\includegraphics[width=7cm]{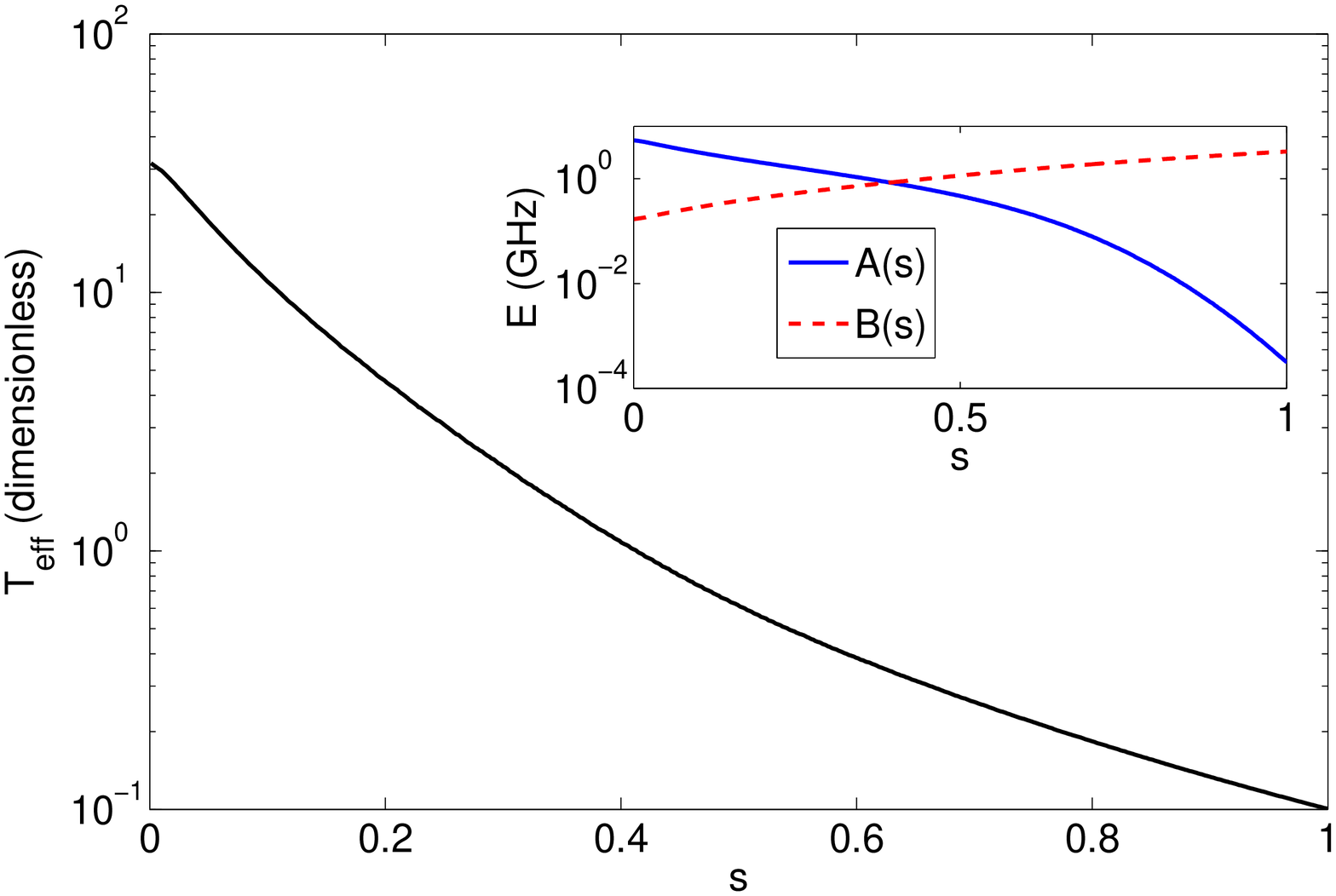}
\par\end{centering}

\caption{\label{fig:Teff}Calculated values of $T_{eff}$ versus $s$ for the
annealing schedule used by a Vesuvius generation D-Wave device. Annealing
schedule is shown in the inset. }
\end{figure}

For parallel tempering, we need to assign an effective temperature
to each value of $s'$, for calculating swap probabilities and determining
the optimal spacing of $s'$ values used. I will demonstrate in Sec.~\ref{sec:Local_search}
that the control parameter $s'$ corresponds to a point in the annealing
schedule $A(s=s')$, $B(s=s')$. Using this fact, an analogous temperature
for any value of $s'$ can be found in the following way. Assume that
a single qubit subject to a longitudinal field of unit strength is
at the point $s'$ in the annealing schedule, at this point the Hamiltonian
will be,

\begin{equation}
H_{1}(s')=-A(s')\,\sigma^{x}+B(s')\,\sigma^{z}.\label{eq:single_qubit}
\end{equation}
This 2x2 Hamiltonian can be diagonalized analytically yielding the
following ratio in the ground state between the basis states in the
classical basis

\begin{equation}
\frac{\psi(1)}{\psi(2)}=\frac{\sqrt{A(s')^{2}+B(s')^{2}}}{A(s')}+\frac{B(s')}{A(s')}.
\end{equation}
By comparing the quantum probability distribution to a Boltzmann distribution
on only the longitudinal part of Eq. \ref{eq:single_qubit} we can
derive an effective temperature,

\begin{equation}
T_{eff}(s')=2\,\left[\ln\left(\left|\frac{\psi(1)}{\psi(2)}\right|^{2}\right)\right]^{-1}.\label{eq:Teff}
\end{equation}
For reasons I explain later, this effective temperature may not necessarily
be a useful approximation of the temperature of the actual observed
raw output of an annealing run, but rather a proxy to establish the
strength of the fluctuations which cause the system to tunnel. As
a demonstration of this technique, Fig.~\ref{fig:Teff} illustrates
an example of calculated $T_{eff}$ for the annealing schedule of
a Vesuvius generation D-Wave device.

Now that we have $T_{eff}(s')$, we can initialize $N$ `replicas',
each with a unique $s_{i}'$, in random states and apply sequential
local searches along with the standard swapping rule for parallel
tempering \cite{Earl2005}, 

\[
P_{swap}(i,j)=
\]

\begin{equation}
\min\left[1,\exp\left(\left(\frac{1}{T_{eff}(s_{i}')}-\frac{1}{T_{eff}(s_{j}')}\right)\left(E_{i}-E_{j}\right)\right)\right]\label{eq:swap_prob}
\end{equation}
where $E_{i}$ is a measure of the energy of each distribution, a
natural choice for $E_{i}$ and $E_{j}$ is the lowest energy found
in an annealing run, which would act as a measure of the depth of
local minima in the potential landscape. This means that the algorithm
will swap to a broad search (higher $T_{eff}$) to escape shallow
local minima, but search in a more focused way (lower $T_{eff}$)
in deep minima. The values of $s'$ which we use for this algorithm
can be chosen based on $T_{eff}(s')$ as discussed in \cite{Earl2005}.

Because $T_{eff}(s')$ is not an actual temperature this analog of
parallel tempering will also not obey detailed balance exactly, even
in the long time limit, however, as I discuss in Sec. \ref{sec:Applications-to-Sampling},
we can still use this algorithm for approximate sampling by performing
some classical post processing. 

In this analogue of parallel tempering we define the ``state'' of
a replica as the lowest energy state found in the previous annealing
run. With this definition, we have all of the ingredients for a complete
parallel tempering analogue, as given in algorithm \ref{alg:parallel_tempering}.
Note that if we use this algorithm with $E$ in Eq. \ref{eq:swap_prob}
defined as the minimum energy found, then we can extract the lowest
energy found in the algorithm where we define a `virtual' $s'=1$
state for which $T_{eff}=+0$ and the annealing protocol is not actually
called, but rather the results simply consist of the state and its
energy. 

\begin{algorithm*}
\begin{algorithmic}[1]
\Procedure{quantum\_parallel\_tempering}{$H,s\_primes,Nstep$} \Comment{Quantum analogue of parallel tempering}
\State $Nqubits \gets length(H)$
\State $states \gets \textrm{RANDOM}\_\textrm{ZEROS}\_\textrm{AND}\_\textrm{ONES}(length(s\_primes),Nqubits)$ \Comment{Initialize random array of states}
\State $energies \gets zeros(length(s\_prime),1)$
\For{$i=1,i++,Nstep$}
	\For{$j=1,j++,length(s\_primes)$} 
		\State $results \gets \textrm{ANNEALER}\_\textrm{CALL}(H,states(j,:),s\_prime)$ \Comment{Call to quantum annealer}
		\State $states(j,:) \gets \textrm{LOWEST}\_\textrm{ENERGY}\_\textrm{STATE}(results)$ \Comment{Extract lowest energy state}
		\State $energies(j) \gets \textrm{MINIMUM}\_\textrm{ENERGY}(results)$ \Comment{May want to use mean energy instead for sampling applications \label{line:min_en}}
	\EndFor
	\State \Return $states$ \Comment{States at each step may be useful for sampling}
	\For{$j=1,j++,length(s\_primes)$}
		\For{$k=j+1,k++,length(s\_primes)$}
			\State $prob \gets \textrm{SWAP}\_\textrm{PROB}(energies(j),energies(k),s\_primes(j),s\_primes(k))$ \Comment Apply Eq. \ref{eq:swap_prob} 
			\If{\textrm{RAND}()<prob} \Comment{Uniform random number between 0 and 1} 
				\State $temp\_state \gets states(j,:)$
				\State $states(j,:) \gets states(k,:)$
				\State $states(k,:) \gets temp\_state$
			\EndIf
		\EndFor
	\EndFor
\EndFor	
\EndProcedure
\end{algorithmic}

\caption{Quantum Analogue of Parallel Tempering, see Sec. \ref{sec:Local_search}
for the details of how the local search given in ANNEALER\_CALL can
be implemented.\label{alg:parallel_tempering}}
\end{algorithm*}

Let us finally consider a different approach, one in which we do runs
with progressively larger $s'$ and therefore progressively smaller
$T_{eff}(s')$. Let us take our inspiration for this from the classical
Monte Carlo technique of population annealing which was introduced
in \cite{Hukushima2003} and further discussed in \cite{Matcha2010,Wang2015}.
In this technique, many replicas are run under SA and at each temperature
step replicas are either destroyed or copied based on a probabilistic
criteria related to their energy. 

Following the classical prescription, for a given replica, the mean
number of copies of that replica which will appear at the next step
will be 

\begin{equation}
\bar{N}(E)=\frac{1}{Q}\exp\left(\left(\frac{1}{T_{eff}(s'_{old})}-\frac{1}{T_{eff}(s'_{new})}\right)E\right),\label{eq:unnorm_prob}
\end{equation}
where 

\begin{equation}
Q=\frac{1}{\bar{N}_{rep}}\sum_{i}\exp\left(\left(\frac{1}{T_{eff}(s'_{old})}-\frac{1}{T_{eff}(s'_{new})}\right)E_{i}\right).\label{eq:normalize}
\end{equation}
The choice of the normalization constant $Q$ guarantees that, while
the total number of replicas will fluctuate from step to step, the
mean number of replicas will remain $\bar{N}_{rep}$ throughout the
algorithm. Without this properly defined normalization, the number
of replicas would either grow exponentially and cause the algorithm
to become impractically slow, or reduce to zero, leaving the outcome
undefined. We now simply follow the same prescription as we did with
parallel tempering, defining the ``state'' of a replica as the minimum
energy found in a given annealing run, and defining the energy of
that replica as the corresponding energy. From this we can now construct
the quantum analogue of population annealing shown in algorithm \ref{alg:population_annealing}.
The effect of this algorithm is to identify and preferentially search
with increasingly short range local searches regions with deep local
minima.

\begin{algorithm*}
\begin{algorithmic}[1]
\Procedure{quantum\_population\_annealing}{$H,s\_primes,Nbar$} \Comment{Quantum analogue of population annealing}
\State $states \gets \textrm{RANDOM}\_\textrm{ZEROS}\_\textrm{AND}\_\textrm{ONES}(Nbar,Nqubits)$ \Comment{Initialize random array of states}
\State $energies \gets zeros(Nbar,1)$
\State $N \gets Nbar$
\For{$i=1,i++,length(s\_primes)-1$}
	\State $newEnergies \gets []$ \Comment{Initialize empty variable}
	\State $newStates \gets []$ \Comment{Initialize empty variable}
	\For{$j=1,j++,N$} 	
		\State $results \gets \textrm{ANNEALER}\_\textrm{CALL}(H,states(j,:),s\_prime)$ \Comment{Call to quantum annealer}
		\State $states(j,:) \gets \textrm{LOWEST}\_\textrm{ENERGY}\_\textrm{STATE}(results)$ \Comment{Extract lowest energy state}
		\State $energies(j) \gets \textrm{MINIMUM}\_\textrm{ENERGY}(results)$ \Comment{May want to use mean energy instead for sampling applications \label{line:min_en_pop}}
	\EndFor
	\State $Q \gets \mathrm{NORMALIZATION}\_\mathrm{FACTOR}(energies,Nbar,s\_primes(i),s\_primes(i+1))$ \Comment{Apply Eq. \ref{eq:normalize}}
	\For{$j=1,j++,N$} 
		\State $Nmean \gets \mathrm{UNNORMALISED}\_\mathrm{PROB}(energies(j),s\_primes(i),s\_primes(i+1))/Q$ \Comment{Apply Eq. \ref{eq:unnorm_prob}}
		\State $Ncopy \gets \mathrm{POISSON}\_\mathrm{RANDOM}\_\mathrm{NUMBER}(Nmean)$ \Comment{Poisson distributed random number}
		\If{Ncopy>0}
			\For{$k=1,k++,Ncopy$}
				\State $\mathrm{APPEND}(newEnergies,energies(j))$ \Comment{Append to list of energies}
				\State $\mathrm{APPEND}(newStates,states(j,:))$ \Comment{Append to list of states}
			\EndFor
		\EndIf
	\EndFor
	\State $N \gets length(newEnergies)$
    \State $energies \gets(newEnergies)$
	\State $states \gets(newStates)$
	\State \Return states 
\EndFor
\EndProcedure
\end{algorithmic}

\caption{Quantum analogue of population annealing see Sec. \ref{sec:Local_search}
for the details of how the local search given in ANNEALER\_CALL can
be implemented.\label{alg:population_annealing}}
\end{algorithm*}

\section{Local Search on an Annealer\label{sec:Local_search}}

For a useful local search we desire two properties, firstly the search
should be local in the sense that it only explores a fraction of the
states in the state space and secondly the search should seek out
more optimal (lower energy) solutions over less optimal ones. Consider
a protocol to search the phase space near a chosen classical state
in the presence of a low temperature bath. The system is first initialized
at $s=1$ in a state which specifies the starting point of the algorithm
and therefore the region to be searched. Local search with a controllable
range is then performed by decreasing the annealing parameter $s$
in Eq. \ref{eq:AnnealerHam} to a prescribed value $s'$ (thereby
turning on a transverse field), possibly waiting for a period of time,
and then returning to $s=1$ and reading out the final state normally.
The low temperature bath will moderate transitions between states,
with detailed balance acting as a guarantee that more optimal states
will be favored in the search. 

One model which has been able to successfully predict experimental
results \cite{Boixo2016,Boixo2014,Dickson2013,Jonson2011,Chancellor2016}
is to assume decoherence acts in the energy eigenbasis. In this model,
which arises from a perturbative expansion in coupling strength \cite{Breuer_book,Albash2014},
coherence can be lost rapidly between energy eigenstates and transitions
between these states can be mediated by the bath but the eigenstates
themselves are not disrupted by the bath. Because the eigenstates
themselves will generally be highly quantum objects, even a completely
incoherent superposition of them can still support quantum effects.

Solving problems using tunneling mediated by open quantum system effects
means that even if the system is initialized in an excited state,
interactions with the environment will cause probability transitions
to other eigenstates. Detailed balance implies that for a bath with
finite temperature the transitions will occur preferentially toward
lower energy states. Furthermore, if $A(s)$ is appropriately small
compared to $B(s)$ in Eq. (\ref{eq:AnnealerHam}) then the quantum
fluctuations can be viewed as local fluctuations around a classical
state, the stronger $A(s)$ is compared to $B(s)$, the less local
this search will be. Consider the perturbative expansion around a
(non-degenerate) classical state $\ket{C(s=1)}$ which can be written
as,

\begin{equation}
\ket{C(s)}=\frac{1}{\mathcal{N}}\sum_{n=0}^{\infty} \left(\frac{A(s)}{B(s)} \right) ^{n}\mathbf{D}_{n}\,\left(\sum_{i}\sigma_{i}^{x}\right)^{n}\ket{C(1)}\label{eq:class_pert}
\end{equation}
where $\mathbf{D}_{n}$ is a diagonal matrix which depends on the
spectrum of $H_{Problem}$ and $\mathcal{N}$ is a normalization factor.
If we assume dephasing noise, then the tunneling rate between two
perturbed classical states, $\ket{C(s)}$ and $\ket{C'(s)}$ will
be proportional to $\sandwich{C(s)}{\sum_{i}\sigma_{i}^{z}}{C'(s)}$.
By inserting the state given in Eq. \ref{eq:class_pert}, we see that
\begin{equation}
\sandwich{C(s)}{\sum_{i}\sigma_{i}^{z}}{C'(s)}\propto \left(\frac{A(s)}{B(s)}\right)^{\mathcal{H}(C(1),C'(1))}+\ldots
\end{equation}
where $\mathcal{H}(C(1),C'(1))$ is the Hamming distance (number of
edges required to traverse on the hypercube) between the two classical
states and $\ldots$ indicates higher powers of $\frac{A(s)}{B(s)}$.
For small $\frac{A(s)}{B(s)}$, tunneling between perturbed classical
states is therefore \emph{exponentially} suppressed in the Hamming
distance between the states. As an eigenstate of a transverse field
Ising model $\ket{C(s\neq1,0)}$ is a fundamentally quantum object which
will exhibit quantum entanglement and therefore will be able to mediate
tunneling between classical states quantum mechanically. We therefore
expect a quantum advantage to be preserved within the local search.
By using quantum searches only locally we have removed the possibility
of gaining a quantum advantage for long range searches beyond the
range of each local search. However, for this price we gain a major
advantage, the classical long range search can be done using state-of-the-art
techniques such as parallel tempering or population annealing, therefore
a small quantum advantage in the local search still results in an
improvement over the underlying classical algorithm. By contrast,
traditional quantum annealing only represents algorithmic improvement
over classical methods if the quantum advantage is \emph{at least}
as large as the advantage which state-of-the-art classical techniques
such as parallel tempering have over simulated annealing.

The question is now how we can program the initial state. The initial
states required for the local search protocol are completely classical,
and therefore could be programmed directly by manipulating the qubits
in a classical way. Another completely classical method would be to
prepare a simple energy landscape where the desired state has the
lowest energy and first heat and then cool the system, thus preparing
it by classical thermal annealing. Both of these methods would require
additional controls or degrees of freedom which may not be accessible
on a real device. For this reason I will instead focus on preparing
the initial state using the standard QAA, which an annealer is able
to perform \emph{by definition}. This is accomplished by running the
QAA with a simple Hamiltonian to guarantee that the system is initialized
in a desired state $y$ ($y(i)\in\left\{ -1,1\right\} $) with a high
probability, for example
\begin{equation}
H_{init}(y)=-\sum_{i}\,y(i)\,\sigma_{i}^{z}-\sum_{i,j\in\chi}y(i)\,y(j)\,\sigma_{i}^{z}\sigma_{j}^{z},\label{eq:Hinit}
\end{equation}
which is a gauge transform of an unfrustrated ferromagnetic system
in a field, and will have a very simple energy landscape and a relatively
large energy difference between the lowest energy and first excited
state. Annealing runs with this Hamiltonian therefore should therefore
reach the target state $y$ with a high probability. After this step,
one needs to be able to reprogram $H_{Problem}$ in Eq. \ref{eq:AnnealerHam}
to be the Hamiltonian of the problem in which we are interested. I
will discuss the feasibility of performing such a protocol on real
annealers in Sec. \ref{sec:Hardware-Implementation}.

\begin{figure}
\begin{centering}
\includegraphics[width=7cm]{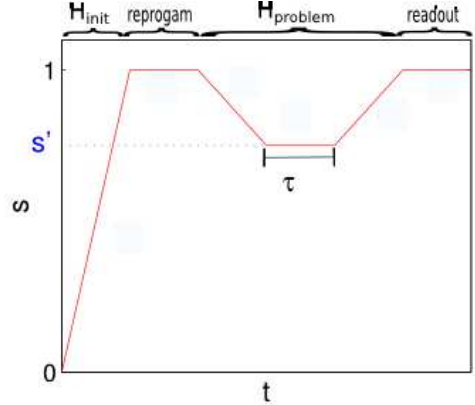}
\par\end{centering}

\caption{\label{fig:runback}Schematic representation of a single annealing
cycle to perform a local search. First the standard QAA is implemented
with the problem Hamiltonian given in Eq. \ref{eq:Hinit} to initialize
the qubits in the desired state. After this the Hamiltonian is reprogrammed
and the device is annealed to $s'$ and optionally allowed to remain
at that point for a time $\tau$. The device is then annealed to $s=1$
and read out. }

\end{figure}

Once we have the desired initial state and problem Hamiltonian programmed,
we simply need to turn on a desired strength of transverse field,
controlled by the value $s'$ shown in Fig.~\ref{fig:runback}. It
may also be desirable to wait for a time $\tau$ before turning the
field off again and reading out the state. The readout does not need
to be any different than what is used with the standard QAA. While
I will not generally assume the capability to anneal to $s=s'$ instantaneously,
as the annealing rate on real devices is sometimes experimentally
limited \cite{Ronnow2014}, it is worth pointing out that in the limit
each run is effectively a noisy continuous time quantum random walk
with a localized starting condition and a finite temperature bath.
In a related work I will examine the relationship between QAA and
quantum random walks in the context of global rather than local searches
\cite{Chancellor_in_prep}.

The function ANNEALER\_CALL in algorithms \ref{alg:adaptive_s}, \ref{alg:parallel_tempering},
and \ref{alg:population_annealing} can be constructed by repeatedly
performing the annealing cycle protocol illustrated in Fig.~\ref{fig:runback}
with the same value of $s'$ and the same initial state $\ket{C(s=0)}$
each time. This function than returns a list of the final state found
in each successive annealing cycle, which is can be thought of as
the results of a probabilistic local quantum search around $\ket{C(s=0)}.$ 
In the next section, I provide numerical demonstrations of the principle of local quantum search.

\section{Proof of Principle\label{sec:proof_of_principle}}

\subsection{methods}

Now that I have discussed how a local search algorithm can be constructed, I will perform some numerical experiments to act as a proof of principle for these methods. While a full numerical examination of all techniques discussed here is beyond the scope of this paper, it is instructive to examine the behaviour in some simple cases to demonstrate that the underlying principles are sound. The numerical method which I will use for this is path integral quantum Monte Carlo, which has previously been used to study the behaviour of quantum annealing \cite{Denchev(2016),Isakov(2016),Santoro(2002),Ronnow(2014),Heim(2015),Martonak(2002)}, this idea is often referred to as Path Integral Quantum Annealing (PIQA). 

PIQA is based on approximating the quantum partition function as a classical partition function of a Hamiltonian consisting of multiple coupled copies of the original Hamiltonian, I do this following the methods of \cite{Martonak(2002)}, using $P=60$ Trotter slices to simulate a system at $T=0.05$, meaning that the actual temperature of the coupled copies is  $P\,T=3$. The number of Monte Carlo steps per Spin (MCS), $\tau_{PIQA}$ is varied in different numerical experiments as explained later. For more details of the numerical methodology, including classical pre-annealing, I refer the reader to the appendix.

While these techniques do not simulate the details of the system bath interactions, they do use local updates to approach a Boltzmann distribution relative to a quantum Hamiltonian. This can roughly be thought of as similar to the action of a bath which performs local updates which obey detailed balance. It is often the case in quantum annealing that tunnelling mediated by these interactions dominates over other factors such as coherence between energy eigenstates, which are not present in PIQA. It has in fact been shown experimentally that similar scaling can be obtained with PIQA as with hardware physical annealers, albeit with a large constant factor advantage in favour of the hardware \cite{Denchev(2016)}.

\begin{figure}
\begin{centering}
\includegraphics[width=7cm]{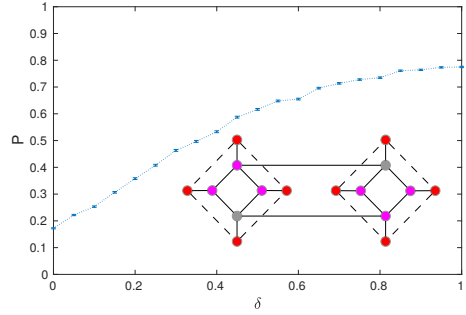}
\par\end{centering}

\caption{\label{fig:therm_asst_ham} Inset: Ising Hamiltonian used in proof of principle studies. Solid lines are ferromagnetic couplers with strength $J=-1$, dashed lines are ferromagnetic couplers with strength $J=-\delta$. Red circles indicate qubits with applied field of $+1$, magenta indicate field of $-1$, and grey indicate no field. Everywhere except for  in the main plot of this figure, I use $\delta=0.2$. Main plot: Simulated annealing success probabilites $P$ with various values of $\delta$ for a sweep with $1000$ steps using a linear anneal starting at $T=10$. Error bars represent one standard deviation of the mean.}

\end{figure}

For these proof of principle studies, I consider a modified version of a Hamiltonian which has been previously used in an experimental quantum annealing study \cite{Dickson2013}. This Hamiltonian has been chosen because it is known to be difficult to solve using  quantum annealing, and was previously used to demonstrate the beneficial role which thermal fluctuations play in solving problems on real world devices. As Fig.~\ref{fig:therm_asst_ham} demonstrates, we have added additional couplers of strength $\delta$ to this Hamiltonian which increase the roughness of the energy landscape. Except for in Fig.~\ref{fig:therm_asst_ham}, I use $\delta=0.2$. I have also observed numerically that these couplers make the problem easier for the PIQA, but standard PIQA still performs relatively poorly in the regime we examine. I have elected to use a linear annealing schedule for these studies $A(s)=1-s$, $B(s)=s$ owing to its relative simplicity. 

This problem is not particularly hard for classical algorithms, as demonstrated by the high success rate of simulated annealing depected in Fig.~\ref{fig:therm_asst_ham}. The relative ease of this problem for classical Monte Carlo approaches is likely due to the small state space of this problem which is constructed of relatively few spins, it can be solved in less than a second by exhaustive search. As the title of this section implies, the calculations here are not intended to prove a scaling advantage of these protocols, but rather a proof of principle of the underlying mechanisms.

One advantage of this Hamiltonian is that it can be implemented on a D-Wave chimera graph, meaning that in principle this Hamiltonian could be used as an experimental tool on real world annealers. However that the values of $T$ and $\tau_{PIQA}$ for the numerical studies here have not been chosen to match the parameters of any of these devices.  

\subsection{results}

\begin{figure}
\begin{centering}
\includegraphics[width=7cm]{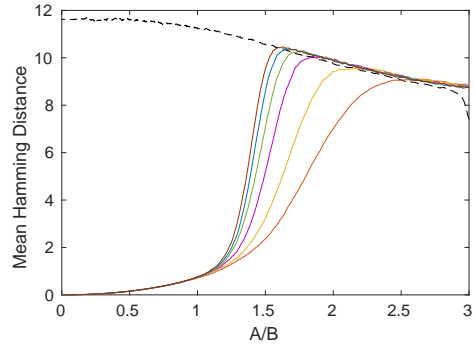}
\par\end{centering}

\caption{\label{fig:runback_dist}Hamming distance from true final ground state versus $A/B$. The simulation is started in the true ground state at $s=1$ and annealing backward using PIQA with $\tau_{PIQA}=500$ (red), $\tau_{PIQA}=1000$ (orange), $\tau_{PIQA}=2000$ (yellow), $\tau_{PIQA}=3000$ (purple), $\tau_{PIQA}=4000$ (green), and $\tau_{PIQA}=5000$ (cyan). The dashed black line is a PIQA run starting at $A(s)/B(s)=3$ and annealing to $s=1$ ($B(s)\gg A(s)$) with a linear schedule and $\tau_{PIQA}=1000$ . All curves are the result of averaging 1,000 individual PIQA runs with $T=0.05$ and $P=60$. Mean Hamming distance is calculated by averaging between all Trotter slices and PIQA runs. The difference between the dashed curve and the solid curves at $A/B=3$ is likely a statistical artifact relating to how the PIQA for the dashed line is initialized, see appendix for details.}

\end{figure}

Let us first examine the effective range explored when annealing hardware is programmed in an initial state at $s=1$ and $s$ is decreased linearly to a value $s'$ at a constant rate. As Fig.~\ref{fig:runback_dist} illustrates, Hamming distance from the initial state, in this case the true ground state of the final Hamiltonian illustrated in Fig.~\ref{fig:therm_asst_ham}. increases continuously as $s$ is decreased until the annealing trajectory merges with that of a traditional PIQA run which starts at small $s$ and then for which $s$ is increased. This plot clearly demonstrates how $s'$ and the annealing rate (which acts as a proxy for wait time at $s'$) may be used to control the typical range of a search.

Let us now consider an example problem to demonstrate the action of the local search mechanism. I first note that none of the $1000$ PIQA runs using a traditional annealing schedule which were used to produce Fig.~\ref{fig:runback_dist} were able to find the correct ground state, demonstrating that this problem is relatively hard for quantum annealing as it is usually formulated. Let us instead consider a very simple hybrid algorithm: initializing in a randomly selected state at $s=1$, anneal to $s'$, wait a period of time $\tau$ and then anneal back. To gain a fair comparison between different values of $s'$, I elect to fix the total number of steps $\tau_{PIQA}=1000$.  

\begin{figure}
\begin{centering}
\includegraphics[width=7cm]{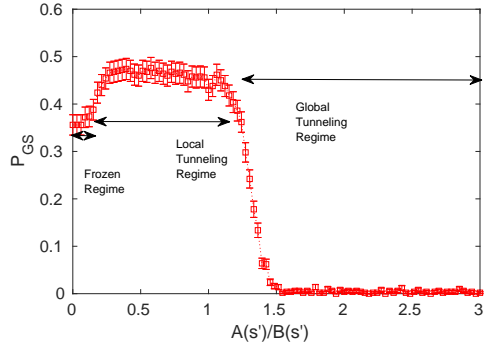}
\par\end{centering}

\caption{\label{fig:runback_perform}Probability of finding the ground state versus $\frac{A(s')}{B(s')}$. These data show three regimes, a frozen regime where no quantum tunneling can occur, a local tunneling regime where performance is enhanced by quantum tunneling from nearby states into the final ground state, and a global tunneling regime, where the state can tunnel throughout the system, and therefore gets trapped in a false minimum with high probability. This figure is based on accumulated statisitics of $500$ individual random starting points which were all subjected to the same classical pre-anneal. Error bars represent one standard deviation of the mean.}

\end{figure}

Fig.~\ref{fig:runback_perform} shows the results of this protocol. Firstly, I note that even at effectively zero transverse field, the system is able to find the final ground state with moderate probability, this is an artifact of the small problem size. For relatively small $\frac{A(s')}{B(s')}$, the probability of finding the ground state remains fixed at this same value, this is the \emph{frozen regime}, where quantum fluctuations are not strong enough to mediate tunneling. At a larger value of $\frac{A(s')}{B(s')}$, the probability of finding the ground state increases due to tunneling from nearby states, it is in this \emph{local tunneling regime} where we can see the advantage of local quantum searching, the state is transfered to a nearby local minimum without becoming trapped in the false minimum which prevents traditionally formulated quantum annealing from performing well. Finally, if $\frac{A(s')}{B(s')}$ is too large, the system tunnels into the false minima and the success rate drops off, this regime, is the \emph{global tunneling regime}. 

This simple example acts as a proof of principle for the power of a local quantum search, specifically, that local quantum exploration can lead to success even when a global search fails. While the problem used in this example is small, and relatively simple, the same principles will hold for larger examples where the rough region is harder to explore classically.

\section{Applications to Sampling\label{sec:Applications-to-Sampling}}

While the parallel tempering-like routine in algorithm \ref{alg:parallel_tempering}
will not obey detailed balance exactly, it may do so in an approximate
sense and therefore may approximate a fair sampling of the local minima
of the energy landscape. Because the system must be annealed to the
point $s=1$ at the end of each call to the annealer, the raw states
extracted from this algorithm will not provide a good approximation
of a thermal sample of the problem Hamiltonian unless this step is
performed very fast. On the other hand, these states do still provide
very approximate information on the relative thermal probabilities
to be found in different local minima, with quantum fluctuations acting
as a rough proxy for thermal fluctuations. It has been shown in \cite{Amin2016}
that for some machine learning tasks a quantum distribution at finite
temperature is actually preferable to a Boltzmann distribution on
a classical Ising model. For other tasks it has been shown that using
quantum fluctuations as a proxy for thermal distributions is sub-optimal,
but makes little difference in practice \cite{Otsubo(2012),Otsubo(2014)}.
An approximate thermal sample can be obtained by first taking all
of the data from the states output by algorithm \ref{alg:parallel_tempering}
at a desired $T_{eff}$ (perhaps throwing the first few iterations
away as warm up period), and then applying the metropolis rules at
$T=T_{eff}$ to obtain a fair thermal sample within each of these
local minima. 

For the estimation of the relative importance of local minima for sampling
purposes, the mean energy found in each run may actually be a more
appropriate measure, rather then the minimum energy which the annealer
found, and therefore more accurate sampling results may be obtained
if the minimum in line \ref{line:min_en} of algorithm \ref{alg:parallel_tempering}
is replaced by the mean of all of the energies found in a run.

Similarly to the parallel tempering analogue, the population annealing
analogue can be used to approximate thermal weight of different
local minima and therefore to sample with appropriate post-processing.
As with the parallel tempering the mean energy in an annealing run
may be a more appropriate choice then the minimum energy as the criterion
for how many copies of a replica to carry on to the next stage of
annealing.

For the more generalized hybrid algorithm, or if an exact Boltzmann
distribution over the classical Ising Hamiltonian is desired, how
to perform sampling is less clear, but sampling could still benefit
from using these algorithms. While the hybrid algorithm may be able
to give a more complete accounting of low energy local minima then
other methods, it does not provide any clue on the relative thermal
weights of each. In principle however, classical post processing could
allow an estimate of the entropy and therefore free energy in disjoint
local minima. By integrating the specific heat numerically using Monte
Carlo techniques, these free energies could then be used find the
appropriate weightings to calculate an overall Boltzmann distribution. 

It is also worth noting that if fast annealing were available, as
suggested in \cite{Amin2016}, then post processing would not be necessary
for either the population annealing or parallel tempering as the distribution
within each disjoint minimum would already be of the form desired
for a quantum Boltzmann machine.

\section{How to avoid solving the wrong problem\label{sec:How-to-avoid}}

Problem mis-specification, where the controls for stating the problem
(the Hamiltonian for the QAA) do not match what the user intended,
is a major difficulty in analog computing \cite{Bissel2004}. In real
devices these mis-specifications come from a variety of sources, such
as low frequency noise from the environment which mimics the control
device or simply from the fact that the controls lack the precision
to represent the actual intended problem. For the purposes of this
discussion the source of the control error does not matter, only that
it is effectively random and relatively uncorrelated. On real devices,
techniques such as gauge averaging can be used to get rid of any correlation
in the control error, and non-random components can be removed by
repeated measurement and adjustments of the device.

The effect of problem mis-specification is that the classical energies
of states are changed randomly. The typical change in energy for any
state in the phase space will be proportional to the energy from applying
a Hamiltonian corresponding to the mis-specifications to that state.
By general statistical arguments this energy shift will be proportional
to $\sqrt{N}$, where $N$ is the total system size \cite{Young(2013)}.
If the mis-specification is strong enough compared to the energy difference
between local versus global energy minima, then the optimum of the
mis-specified Hamiltonian will no longer correspond to the optimal
solution of the target problem, and even a machine which finds the
lowest energy state perfectly on the mis-specified Hamiltonian can
only obtain an approximate solution.

Problem mis-specification means that, as a device to perform the QAA
is scaled, the errors must also be reduced or the device will no longer
be able to reliably optimize, and any fundamental limit on the precision
of the controls will also be a limit on the size of problem for which
a such a device will be useful. The reason for this is that the QAA
is a global search, corruption of the energy landscape anywhere in
the phase space has the potential to destroy performance. 

For a local search however, the total size of the space is irrelevant
to the performance of the search. The reason is as follows, which
state is optimum depends only on the relative energy of states which
are explored, clearly shifts relative to the energy of states which
are not searched will have no effect. Errors due to problem mis-specification
arise from the device being effectively `tricked' by bitstrings which
have falsely been assigned a lower energy than the actual solution.
The deviation in energy differences between the searched bitstrings
is therefore the relevant quantity, and overall energy shifts in all
states which can be reached are irrelevant. 

Local searches are not immune to problem mis-specification. However,
if I assume that the shape of the subspace explored is itself roughly
in the shape of a hypercube then by the same arguments as \cite{Young(2013)},
the relevant energy shift caused by the problem being incorrectly
specified will go as $\sqrt{\mathfrak{h}(s')}$ where $\mathfrak{h}(s')$
is the typical Hamming distance between the states explored. While
$\mathfrak{h}(s')$ itself may scale with $N$ for a fixed $s'$,
the value of $s'$ can be adjusted for different system size based
on the algorithm \ref{alg:adaptive_s}, and therefore held fixed,
thus removing the direct dependence of the error on $N$. As I discussed
previously, fixing the range of the search does mean that a quantum
advantage cannot be obtained beyond the range of the search, however,
it does not preclude the possibility of such an advantage within the
search range. 

Even if the assumption that the subspace explored by the local search
is roughly hypercubic in shape is relaxed, then the effective typical
energy shift can still be upper-bounded using the following arguments.
Increasing the number of states in a search space can never decrease
the probability of the search space being corrupted by problem mis-specification.
This is intuitively clear because if a bitstring exists in the search
space which due to problem mis-specification has a lower energy than
the correct solution, this bitstring will still be in the search space
if it is expanded, and the search space will still be corrupted. Therefore,
if the furthest Hamming distance between states reached in a local
search is $\mathfrak{h}_{max}(s')$ than the probability of state
space corruption for the search space has to be less than the probability
of corruption in a hypercubic space with size $\mathfrak{h}_{max}(s')$
and therefore the effective typical energy shift must be less than
$\sqrt{\mathfrak{h}_{max}(s')}$, which again could be substantially
less then the total system size, $N$.

In principle local searches are still a valid optimization technique
for arbitrarily large Hamiltonians as long as each search does not
search too large a subspace. 

\section{Hardware Implementation\label{sec:Hardware-Implementation}}

Let us now examine the question of whether the reprogramming step
in Fig.~\ref{fig:runback} is experimentally feasible. To do this
we must discuss the details of how an annealer actually works. In
this section I will first discuss in general terms why the methods
proposed here should be practical on any annealer, this will be followed
by a brief discussion of the feasibility of doing such searches specifically
on the superconducting circuit hardware constructed by D-Wave Systems
Inc. 

The local search protocol laid out in this paper does not require
any modification of the parts of the protocol for which quantum mechanics
plays a role. The only additional capability beyond what one should
expect for any useful quantum annealer (programmable Hamiltonians,
ability to adjust parameters over an appropriate range, some degree
of coherent quantum interaction, etc...), is the ability to program
the state of the machine at the end of the anneal. As I have mentioned
previously however, at this point the annealer is in a completely
classical state and there is no need to protect a delicate quantum
superposition. 

If the annealer architecture also supports individually addressable
manipulation of individual qubits by either quantum or classical means
then the necessary state can be initialized by first running the annealer
with the desired Hamiltonian and than measuring the result and applying
appropriate gates to reach the desired state. In this case the problem
Hamiltonian does not ever need to be changed. One example for which
this will certainly be true is a simulation of a quantum annealer
which is run on a universal quantum computer, in this case no initial
annealing run is even necessary, the bits can just be initialized
in the $\ket{0000....}$ state and than appropriate bit flips could
be applied. 

As I have discussed in Sec. \ref{sec:Local_search}, direct manipulation
of the qubits is not necessary, as long as the Hamiltonian can be
reprogrammed at $s=1$ without disturbing the (classical) state of
the bits. As the Hamiltonian must already be programmable for an annealer
to be useful, the necessary criteria for this to be possible is that
the qubits either are naturally stable enough or can be made stable
enough that their state is not changed by reprogramming the Hamiltonian.
Even in a system which lacks natural stability, such stability could
be achieved for example by turning on strong field terms while the
couplers are reprogrammed, and then turning the field terms off again.

\begin{figure}
\begin{centering}
\includegraphics[width=7cm]{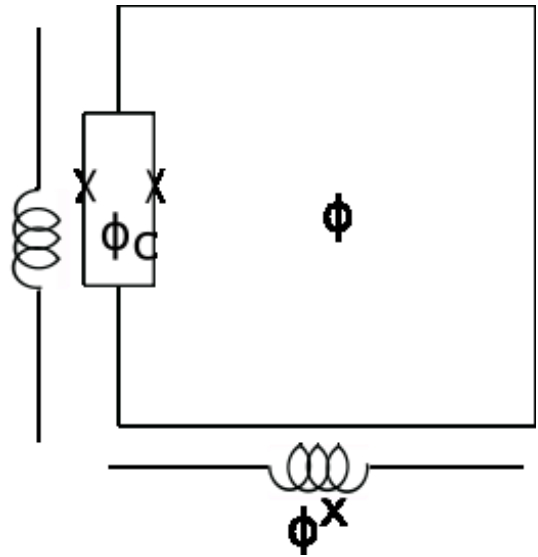}
\par\end{centering}

\caption{Circuit schematic where fluxes through small and large loops control
the effective potential which the flux $\phi$ experiences are illustrated
in Fig.~\ref{fig:potential_cartoon}. For more details about the circuit
design, see \cite{Harris2010}. \label{fig:D-Wave_circuit}}

\end{figure}

I now specifically discuss the superconducting circuit architecture
by D-Wave systems Inc. These devices are based on superconducting
circuits of the type shown schematically in Fig.~\ref{fig:D-Wave_circuit}
\cite{Harris2010}. 

One crucial aspect to the feasibility of the D-Wave hardware for local
searches is that it has been shown experimentally that open quantum
system effects play a major role in the ability of the D-Wave devices
to perform tunneling and find ground states \cite{Dickson2013}. It
is worth pointing out that the importance of open quantum system effects
does \emph{not} indicate that the D-Wave devices are classical rather
than quantum solvers. In fact, for these devices there is experimental
evidence of both entanglement \cite{Lanting2014} and tunneling \cite{Boixo2016,Boixo2014}. 

\begin{figure}
\begin{centering}
\includegraphics[width=7cm]{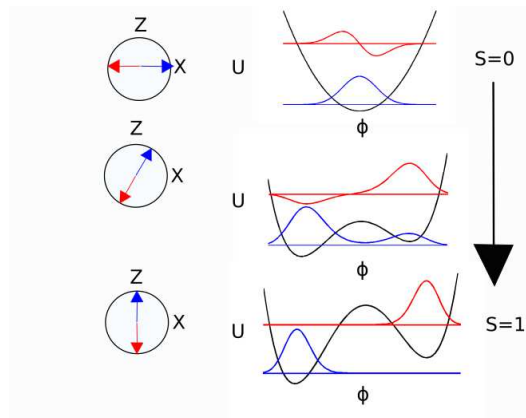}
\par\end{centering}

\caption{\label{fig:potential_cartoon}Schematic figure illustrating how the
D-Wave annealer implements the Ising Model. left: Representation of
ground and first excited state in the x-z plane of the Bloch sphere.
right: Corresponding potential with ground and first excited states
overlayed. Top: monostable potential corresponding to strong transverse
field (s=0) Mid: bistable potential with quantum tunneling corresponding
to transverse and longitudinal field of similar strength (intermediate
value of s). Bottom: bistable potential where tunneling is no longer
supported corresponding to strong longitudinal field with no transverse
field (s=1). }
\end{figure}

In the D-Wave architecture there is classical superconducting control
circuitry \cite{Johnson2010} which is used to set the fields and
couplers in Eq. \ref{eq:ISGham}. Tuning the components of this circuitry
which are used to program these states will produce stray fields and
heat which one may be concerned could disrupt the state of the qubits.
The final classical state of these qubits should be very stable, and
need not support any delicate quantum superpositions since it is a
classical basis state. The reason for the stability is the way in
which the qubits are implemented. The effective Ising model in the
D-Wave device is produced by circuits which actually implement an
effective double well potential \cite{Jonson2011}, this potential
is then tuned between a mono- and bi- stable regime. The ground and
first excited state of each of these individual wells can be mapped
to an effective Ising spin, for which quantum tunneling between the
wells mediates superposition states as illustrated schematically in
Fig.~\ref{fig:potential_cartoon}. The strength of tunneling depends
exponentially on the width and height of the barriers between these
two wells, and therefore tuning far into the bistable regime strongly
suppresses any quantum tunneling, and makes the qubits effectively
classical. In this classical regime it may still be possible for external
heat to cause the qubit to be excited over the energy barrier thermally.
To avoid this problem one should be able to simply bias the qubits
further into the bistable regime until the barrier heights are sufficient,
or to perform the reprogramming more slowly so heat has more time
to dissipate.

An outstanding question is whether the D-Wave chip as it is currently
constructed is capable of implementing such a protocol. While it has
not been tested experimentally whether or not such reprogramming will
disrupt the state of the qubits on these devices, the low level controls
do exist to perform the protocol in Fig.~\ref{fig:runback} \cite{Lanting_pc}.
If a high fidelity could be obtained in the reprogramming step on
the current device, then the algorithms given in this paper could
be tested by someone with low level access to these devices without
performing any physical modifications. If this is indeed the case,
such experiments could provide a proof-of-principle for the ideas
stated here. To go beyond proof-of-principle it would probably be
desirable to optimize a machine to perform the protocol given in Fig.~\ref{fig:runback}
so that the time to reprogram, and the energy released
in the reprogramming step (which would determine the time needed to
cool afterward) are both minimized.

Even if the suggested reprogramming routine is not possible at high
fidelity on current D-Wave devices, it is likely that it could be
made feasible through design alterations on updated devices, or on
other hardware for simulating transverse field Ising models.

\section{Conclusions\label{sec:Conclusions}}

We have discussed a new protocol using quantum annealers which may
be feasible on current devices with no alterations, 
as well as performed a proof-of-principle numerical experiment to 
demonstrate the idea of a local search using a quantum annealer. I discuss numerous
advantages over the QAA as it is implemented currently. In contrast
to the QAA, the idea I propose can perform a local search rather then
a global search. One advantage of such a search is that the effect
of problem mis-specification depends on the range of the search, rather
then the overall number of qubits, and therefore this method should
work even when noise would ruin a global search with the QAA. I further
argue that this type of search can be used to construct hybrid algorithms,
where quantum and classical searches can be used sequentially to gain
the complementary advantages of each. I argue that an advantage could
be obtained in this case even if the classical algorithm is simply
to randomly initialize the state of the annealer. I furthermore construct
analogues of powerful classical algorithms, but using a quantum processor,
and discuss how these methods can be applied to sampling with appropriate
post-processing. Because the protocol I propose is able to take advantage
of state-of-the-art classical techniques, it will represent an algorithmic
improvement even if the local searches gain only a small quantum advantage,
in contrast the traditional QAA only provides such an improvement
if the quantum advantage is at least as large as the classical advantage
which advanced techniques such as parallel tempering have over simulated
annealing. I have further demonstrated the underlying principles of these algorithms with simple numerical experiments.

\section*{Acknowledgments}

The author was supported by EPSRC (grant ref: EP/L022303/1), and
would like to thank Viv Kendon for several critical readings of the
paper and useful discussions. The author further thanks Trevor Lanting,
Helmut Katzgraber, Gabriel Aeppli, Andrew G. Green, and Paul A. Warburton
for useful discussions.
\section*{Appendix: Path Integral Quantum Annealing}

To perform Path Integral Quantum Annealing (PIQA), I followed the procedures given in \cite{Martonak(2002)}. While I will not reproduce the derivations of that paper, I will touch on several important similarities and differences which are necessary to be aware of to reproduce my numercal results. 

\subsection*{Time Parameter $\tau_{PIQA}$}

The time parameter $\tau_{PIQA}$ counts the number of Monte Carlo steps per spin (MCS). As was done in \cite{Martonak(2002)}, at each time step, a `classical' cluster update was also attempted for each of the spins in the Ising system. These updates consisted of flipping the same spin simultaneously in all Trotter slices.

\subsection*{Starting Conditions for Traditional QAA}

For runs performed with the traditional QAA, I started with $s'$ such that $A/B=3$ rather than at $B=0$. This is to prevent the Trotter slices from becoming completely uncorrelated, and is consistant with the methodology of \cite{Martonak(2002)}.

\subsection*{Classical Pre-anneal}

As was done in \cite{Martonak(2002)}, I initialized all PIQA runs with a classical pre-anneal to achieve the appropriate initial state. This pre-anneal consisted of $100$ Monte Carlo sweeps run on the classical Hamiltonian. After the pre-anneal, all Trotter slices were initialized in this state, the breaking of this uniformity in initialization is likely the cause of the numerical artefact seen near $A/B=3$ for the dashed line in Fig.~\ref{fig:runback_dist}. For runs initialized at $A/B=3$ this classical pre-anneal was performed at $T_{class}=P\,T$, on the other hand for runs starting with $B\gg A$ this was done by taking $T_{class}=T$. 

In cases where multiple PIQA runs are plotted starting at $s=1$, the initial states for the PIQA is taken as the result of a single classical pre-anneal with the same set of initial states. This removes unimportant statistical noise due to the pre-anneal finding different classical states in different cases.  The actual \emph{relative} statistical variations between points in Fig.~\ref{fig:runback_perform} is probably therefore actually substantially smaller than the errorbars, which represent \emph{absolute} statistical variation in the values, although it is difficult to estimate precisely by how much. \\

\end{document}